\documentclass{amsart}
\usepackage{graphicx,tikz,bm}
\usepackage{amsmath,amsfonts,amssymb}
\usepackage{color}
\usetikzlibrary{decorations.markings}
\usepackage[americanresistors,americaninductors]{circuitikz}
\usepackage{steinmetz}
\tikzstyle{vertex}=[circle, draw, inner sep=0pt, minimum size=4pt]
\tikzstyle{bvertex}=[circle, draw, fill=black, inner sep=0pt, minimum size=6pt]
\tikzstyle{emptynode}=[circle]

\newtheorem{theorem}{Theorem}[section]

\theoremstyle{definition}

\newtheorem{assumption}[theorem]{Assumption}

\theoremstyle{remark}
\newtheorem{remark}[theorem]{Remark}
\numberwithin{equation}{section}

\begin{document}
\title{Compositional Transient Stability Analysis\\of Multi-Machine Power Networks} 

\author{Sina Y. Caliskan}
\address{ Department of Electrical Engineering, University of California at Los Angeles, CA 90095-1594, United States}
\email{caliskan@ee.ucla.edu}

\author{Paulo Tabuada}
\address{ Department of Electrical Engineering, University of California at Los Angeles, CA 90095-1594, United States}
\email{tabuada@ee.ucla.edu}


\date{}

\begin{abstract}                        
During the normal operation of a power system all the voltages and currents are sinusoids with a frequency of 60 Hz in America and parts of Asia, or of 50Hz in the rest of the world. Forcing all the currents and voltages to be sinusoids with the right frequency is one of the most important problems in power systems. This problem is known as the transient stability problem in the power systems literature.

The classical models used to study transient stability are based on several implicit assumptions that are violated when transients occur. One such assumption is the use of phasors to study transients. While phasors require sinusoidal waveforms to be well defined, there is no guarantee that waveforms will remain sinusoidal during transients. In this paper, we use energy-based models derived from first principles that are not subject to hard-to-justify classical assumptions. In addition to eliminate assumptions that are known not to hold during transient stages, we derive intuitive conditions ensuring the transient stability of power systems with lossy transmission lines. Furthermore, the conditions for transient stability are compositional in the sense that one infers transient stability of a large power system by checking simple conditions for individual generators.
\end{abstract}

\maketitle

\markboth{COMPOSITIONAL ANALYSIS OF MULTI-MACHINE POWER NETWORKS}{SINA Y. CALISKAN AND PAULO TABUADA}
\section{Introduction}
\label{intro}

Power system is the name given to a collection of devices that generate, transmit, and distribute energy to consuming units such as residential buildings, factories, and street lighting. Abusing language, we use the terms power and energy interchangeably, as typically done in the power systems literature. Excluding a small portion of generating units, such as solar cells and fuel cells, we can think of power generators in a power system as electromechanical systems \cite{BergenBook, SauerBook}. Natural sources, such as the chemical energy trapped in fossil fuels, are used to generate mechanical energy, which is then converted into electrical energy.

When power systems are working in normal operating conditions, i.e., in \textit{steady-state}, the generators satisfy two main conditions: $1)$ their rotors rotate with the same velocity, which is also known as \textit{synchronous velocity}, and $2)$ the generated voltages are sinusoidal waveforms with the same frequency. Keeping the velocity of the generators at the synchronous velocity and the terminal voltages at the desired levels is called \textit{frequency stability} and \textit{voltage stability}, respectively \cite{IEEE2004}. When all the generators are rotating with the same velocity, they are synchronized and the relative differences between the rotor angles remain constant. The ability of a power system to recover and maintain this synchronism is called \textit{rotor angle stability}. \textit{Transient stability}, as defined in \cite{IEEE2004}, is the maintenance of rotor angle stability when the power system is subject to large disturbances. These large disturbances are caused by faults on the power system such as the tripping of a transmission line.

In industry, the most common way of checking transient stability of a power system is to run extensive time-domain simulations for important fault scenarios \cite{PavellaBook}. This way of developing action plans for the maintenance of transient stability is easy and practical \textit{if} we know all the ``important" scenarios that we need to consider. Unfortunately, power systems are large-scale systems and the number of possible scenarios is quite large. Since an exhaustive search of all of these scenarios is impossible, power engineers need to guess the important cases that they need to analyze. These guesses, as made by humans, are prone to errors. Moreover, time-domain simulations do not provide insight for developing control laws that guarantee transient stability \cite{PavellaBook2}. Because of these reasons, additional methods are required for transient stability analysis. Currently, the methods that do not rely on time-domain simulations can be collected in two different groups: direct methods and automatic learning approaches. The latter, automatic learning approaches \cite{Wehenkel1998}, are based on machine learning techniques. In this work, we do not consider automatic learning approaches and we focus on direct methods. 

\subsection{Direct Methods and Their Limitations}

Direct methods are based on obtaining Lyapunov functions for simple models of power systems \cite{FouadBook, PavellaBook, Varaiya1985}. To the best of our knowledge the origin of the idea can be found in the 1947 paper of Magnusson \cite{Magnusson} which uses the concept of ``transient energy" which is the sum of kinetic and potential energies to study the stability of power systems. In 1958, Aylett, assuming that a two-machine system can be represented by the dynamical equation
$$
\ddot{\theta} = B - \sin(\theta),
$$
showed that there exists a separatrix dividing the two-dimensional plane of $\dot{\theta}$ and $\theta$ into two regions \cite{Aylett1958}. One of the regions is an invariant set with respect to the two-machine system dynamics, i.e., if the initial condition is in this set, trajectories stay inside this set for all future time. Aylett concluded that in order to check the stability of the system, we only need to check whether the state is in the invariant set or not. Aylett also characterized the separatrix that defines the invariant set and extended the results from the two-machine case to the three-machine case in the same monograph. Although the term ``Lyapunov function" was not stated explicitly in his work, Aylett's work used Lyapunov-based ideas. Some of the other pioneering works on direct methods include Szendy \cite{Szendy1962}, Gless \cite{Gless1966}, El-Abiad and Nagaphan \cite{ElAbiad1966}, and Willems \cite{Willems1971}. The work based on direct methods mainly focused on finding better Lyapunov functions that work for more detailed models and provide less conservative results. These Lyapunov functions are used to estimate the region of attraction of the stable equilibrium points that correspond to desired operating conditions. The stability of a power system after the clearance of a fault can then be tested by determining if the post-fault state belongs to the desired region of attraction. For further information we refer the reader to \cite{FouadBook, PavellaBook, PavellaBook2, PaiBook, Varaiya1985}. There are several problematic issues with direct methods. 

The first problem is the set of assumptions used to construct these models. The models used for transient stability analysis implicitly assume that the angular velocities of the generators are very close to the synchronous velocity. In other words, it is assumed that the system is very close to desired equilibrium and the models developed based on this assumption are used to analyze the stability of the same equilibrium. The standard answer given to this objection is the following: the models that are used in transient stability studies are used only for the ``first swing" transients and for these transients the angular velocities of the generators are very close to the synchronous velocity. Unfortunately, in real world scenarios large swings need to be considered. Citing the post mortem report~\cite[page 25]{USDOE2003} on the August 14, 2003 blackout in Canada and the Northeast of the United States, ``the
large frequency swings that were induced became
a principal means by which the blackout spread
across a wide area''. Using models based on ``first swing'' assumptions to analyze cases like the August 14, 2003 blackout does not seem reasonable.

The second problem is that the models used for transient stability analysis, again implicitly, pose certain assumptions on the grid. The transmission lines are modeled as impedances and the loads are either modeled as impedances or as constant current sources. These modeling assumptions are used to eliminate the  internal nodes of the network via a procedure called Kron reduction \cite{BergenBook, Dorfler2013}. The resulting network after Kron reduction is a strongly connected network. Every generator is connected to every other generator via transmission lines modeled as a series connection of an inductor and a resistor. After this reduction process, the resistances in the reduced grid are neglected. The fundamental reason behind the neglect of the resistances lies in the strong belief, in the power systems community, about the non-existence of Lyapunov functions when these resistances are not neglected. This belief stems from the paper \cite{Chiang1989} which asserts the non-existence of global Lyapunov functions for power systems with losses in the reduced power grid model. It is further supported by the fact that the Lyapunov functions that the power systems community has developed contain path-dependent terms unless these resistances are neglected. The reader should note that the resistors here represent both the losses on the transmission lines \textit{and} the loads. Hence this assumption implies that there is no load in the grid (other than the loads modeled as current injections), which is not a reasonable assumption. In addition to these problems that have their origin in neglecting the resistances on the grid, the process of constructing these reduced models, i.e. Kron reduction, can only be performed for a very restrictive class of circuits \textit{unless} we assume that all the waveforms in the grid are sinusoidal \cite{Tabuada2012, Tabuada2013}. In other words, in order to perform this reduction process for arbitrary networks, we need to use phasors, which in turn requires that all the waveforms in the grid are sinusoidals and every generator in the power grid is rotating with the same velocity.  This assumption is not compatible with the study of transients.

\subsection{Control and Synchronization in Power Networks}

Despite the long efforts to obtain control laws for power systems with non-negligible transfer conductances, results only appeared in the beginning of the $21^\text{st}$ century. For the single machine and the two machine cases, a solution, under restrictive assumptions, is provided in \cite{Ortega2005}. In the same work, the existence of globally asymptotically stabilizing controllers for power systems with more than two machines is also proved but no explicit controller is suggested. An extension of the results in \cite{Ortega2005} to structure preserving models can be found in \cite{Dib2009}. To the best of our knowledge, the problem of finding explicit globally asymptotically stabilizing controllers for power systems with non-negligible transfer conductances and more than two generators has only been recently solved in \cite{Casagrande2011, Casagrande2012}. Although a solution has been offered for an important long-lasting problem, the models that are used in \cite{Casagrande2012} are still the traditional models that we want to avoid in our work.

There are also some recent related results on synchronization of Kuramoto oscillators \cite{Dorfler2011, Dorfler2012, Dorfler2013-2}. If the generators are taken to be \textit{strongly overdamped}, these synchronization results can be used to analyze the synchronization of power networks. The synchronization conditions obtained in \cite{Dorfler2011, Dorfler2012, Dorfler2013-2} can also be used in certain micro-grid scenarios \cite{Simpson2013}. In this paper, we provide results that do not require generators to be strongly overdamped. 

\subsection{A New Framework for Transient Stability of Power Systems}

All the previously described methods use classical models for power systems. They are only valid when the generator velocities are very close to the synchronous velocity. In this paper, we abandon these models and use port-Hamiltonian systems~\cite{ArjanBook} to model power systems from first principles. As already suggested in~\cite{Arjan2012}, a power system can be represented as the interconnection of individual port-Hamiltonian systems. There are several advantages of this approach. First of all, we have a clear understanding of how energy is moving between components. Secondly, we do not need to use phasors. Thirdly, we do not need to assume all the generator velocities to be close to the synchronous velocity. Finally, using the properties of port-Hamiltonian systems, we can easily obtain the Hamiltonian of the interconnected system, which is a natural candidate for a Lyapunov function. A similar framework, based on passivity, is being used in a research project on the synchronization of oscillators with applications to networks of high-power electronic inverters \cite{Torres2013}.

We first obtain transient stability conditions for generators in isolation from a power system. These conditions show that as long as we have enough dissipation, there will be no loss of synchronization. In \cite{Arjan2012} the port Hamiltonian framework is also used to derive sufficient conditions for the stability of a single generator. The techniques used in \cite{Arjan2012} rely on certain integrability assumptions that require the stator winding resistance to be zero. In contrast, our results hold for non-zero stator resistances. Moreover, while in \cite{Arjan2012} it is assumed that synchronous generators have a single equilibrium, we show in this paper that generators have, in general, 3 equilibria and offer necessary and sufficient conditions on the generator parameters for the existence of a single equilibrium. With the help of useful properties of port-Hamiltonian systems, we obtain sufficient conditions for the transient stability of the interconnected power system from the individual transient stability conditions for the generators. In addition to these sufficient conditions, which were also reported in \cite{Tabuada2013-2}, we provide a deeper discussion on the modeling of synchronous generators and we also explain how to relax the sufficient conditions with the help of FACTS devices. Our results are important contributions for several reasons. Firstly, we do not use the previously discussed questionable assumptions. Without these assumptions, we can apply our conditions to realistic scenarios including cases with large frequency swings. Secondly, our results relate dissipation with transient stability. This transparent relation is hard to see in the classical framework due to shadowing assumptions. Thirdly, we exploit compositionality to tame the complexity of analyzing large-scale systems. We propose simple conditions that can be independently checked for each generator without the need to construct a dynamical model for the whole power system. Finally, extending our framework to more complex models is easier because we use port-Hamiltonian models for the individual components. This flexibility will be helpful to design future control laws for the generators.

\section{Notations}
We denote the diagonal matrix with diagonal elements $d_1,\hdots,d_n$ by \linebreak \mbox{$\textbf{diag}(d_1,\ldots,d_n)$} and an $n$ by $m$ matrix of zeros by $0_{n \times m}$. The vector $v \in \mathbb{R}^n$ is denoted by $(v_1,\ldots,v_n)$, where $v_i$ is its $i^\text{th}$ element. The $n$ by $n$ identity matrix is denoted by $1_n$. We say that $P \in \mathbb{R}^{n \times n}$ is positive semidefinite, denoted by $P \ge 0$, if $x^TPx\ge 0$ for all $x\in \mathbb{R}^n$. If, in addition to this, we also have $x^TPx=0$ only if $x=0$, we call $P$ positive definite, denoted by $P>0$. A matrix $P$ is negative semidefinite (definite), denoted by $P \le 0$ ($P<0$), if and only if $-P$ is positive semidefinite (definite). The gradient of a scalar field $y$ with respect to a vector $x = (x_1,\ldots,x_n)$ is given by
$
\frac{\partial y}{\partial x} = \begin{bmatrix}\frac{\partial y}{\partial x_1} & \ldots & \frac{\partial y}{\partial x_n}\end{bmatrix}^T.
$
Note that the gradient is assumed to be a \textit{column} vector. Consider the affine control system
\begin{equation}
\dot{x} = f(x)+g(x)u\label{eqn:notationds}
\end{equation}
where $x \in \mathcal{M}$, $u \in \mathcal{U}$, $\mathcal{M}$ is a manifold and \mbox{$\mathcal{U}\subseteq \mathbb{R}^m$} is a compact set. The affine control system \eqref{eqn:notationds} has a port-Hamiltonian representation if there exist smooth functions \mbox{$\mathcal{J}:\mathcal{M}\to \mathbb{R}^{n\times n}$} and \mbox{$\mathcal{R}:\mathcal{M}\to \mathbb{R}^{n\times n}$} satisfying
\mbox{$\mathcal{J}^T(x) = -\mathcal{J}(x)$} and \mbox{$\mathcal{R}(x) = \mathcal{R}^T(x) \ge 0$} for all $x \in \mathcal{M}$, and there exists a smooth function \linebreak \mbox{$H:\mathcal{M} \rightarrow \mathbb{R}$}, which is called the Hamiltonian, such that \eqref{eqn:notationds} can be written in the form
\begin{equation}
\dot{x} = (\mathcal{J}(x) - \mathcal{R}(x))\frac{\partial H}{\partial x} + g(x)u.\label{eqn:notationph}
\end{equation}
The Hamiltonian $H$ can be thought of as the total energy of the system. The output of the port-Hamiltonian representation of \eqref{eqn:notationds} is given by
$$y = g^T(x)\frac{\partial H}{\partial x}.$$
If we take the time derivative of the Hamiltonian, we obtain
\begin{equation}
\frac{dH}{dt} = \frac{\partial H}{\partial x}^T\dot{x} = -\frac{\partial H}{\partial x}^T\mathcal{R}\frac{\partial H}{\partial x} + u^Ty \le u^Ty. \label{eqn:passivity}
\end{equation}
The term $u^Ty$ in \eqref{eqn:passivity} represents the power supplied to the system. Therefore, property \eqref{eqn:passivity} states that the rate of increase of the Hamiltonian is less than the power supplied to the system. We refer to \cite{ArjanBook} for further details on port-Hamiltonian systems.

\section{Single Generator}
\label{sec:sg}
In this section, we derive the equations of motion for a two-pole synchronous generator from first principles. The first step in this derivation is to identify the Hamiltonian, the sum of the kinetic and the potential energy, of a single generator. We then derive a stability condition, using the Hamiltonian as a Lyapunov function, for the synchronous generator when the terminal voltages are known. Although we only consider two-pole synchronous machines, the results in this section can easily be generalized to machines with more than two poles.
\subsection{Mechanical Model}
\label{sec:sg:mechanical}
Every synchronous generator consists of two parts: rotor and stator. Several torques act on the rotor shaft and cause the rotor to rotate around its axis. Explicitly, we can write the torque balance equation for the torques acting on the rotor shaft as follows:
\begin{equation}
\label{eqn:torquebalance}
M\ddot{\theta} + D\dot{\theta} = \tau_m - \tau_e,
\end{equation}
where $\theta$ is the rotor angle, $M$ is the moment of inertia of the rotor shaft, $D$ is the damping coefficient, $\tau_m$ is the applied mechanical torque and $\tau_e$ is the electrical torque. The angular velocity of the rotor shaft is $\omega = \dot{\theta}$.  The total kinetic energy of the rotor can be expressed as
$$
H_{\text{kinetic}} = \frac{1}{2}M\omega^2.
$$
Using the definition of the angular velocity $\omega$, we can write \eqref{eqn:torquebalance} in the form
\begin{align}
\dot{\theta} &= \omega,\label{eqn:mech1} \\
M\dot{\omega} &= -D\omega + \tau_m - \tau_e. \label{eqn:mech2}
\end{align}
\begin{remark}
In the classical power systems literature, the torque balance equation \eqref{eqn:torquebalance} is scaled by $\omega$. Defining $P_m = \omega \tau_m$, $P_e = \omega \tau_e$, $M' = M \omega$, $D' = D \omega$ and dividing both sides of \eqref{eqn:torquebalance} by a constant value called rated power, the following set of mechanical equations is obtained:
\begin{align}
\dot{\theta} &= \omega, \label{eqn:mech3} \\
M'\dot{\omega} &= -D'\omega + P_m - P_e. \label{eqn:mech4}
\end{align}
In these equations, the parameters $M'$ and $D'$ are assumed to be constant, which implies that $\omega$ is either constant or \textit{slowly} changing. Equations~(\ref{eqn:mech1}) and~(\ref{eqn:mech2}) do not require such assumptions on $\omega$.
\end{remark}
\subsection{Electrical Model}
\label{sec:sg:electrical}
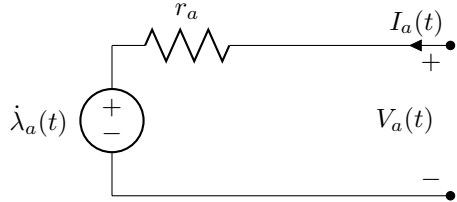
\begin{figure}[h!]
\begin{circuitikz}[american voltages, scale=1, transform shape]
\centering
\draw
  (4.5,0) to [short, *-] (0,0)
  to [V, l=$\dot{\lambda}_{a}(t)$] (0,2)
  to [R, l=$r_a$] (2,2)
  (2,2) to [short, -*] (4.5,2)
  (4.5,0) to [open, v^>=$V_{a}(t)$] (4.5,2)
  (4.1, 2)to [short, i_=$I_{a}(t)$] (4,2);
\end{circuitikz}
\caption{Phase-$a$ stator winding}
\label{phaseafig}
\end{figure}

%In this section, we derive the electrical model for the synchronous generator by applying Kirchoff's Voltage Law to the electrical circuits connected to the rotor and the stator. Magnetic interactions between these circuits couple the equations. We also provide the equations that explicitly describe how this coupling occurs. 
%
There are three identical circuits connected to the stator. These circuits are called \textit{stator windings} and they are labeled with letters $a$, $b$ and $c$. There are also windings connected to the rotor. These winding are called \textit{field windings}. In this work, we consider a synchronous generator with a single field winding. In a cylindrical rotor synchronous generator, which are predominantly used in nuclear and thermal generation units, the aggregated effect of the field windings can be modeled by a single circuit \cite{FitzgeraldBook}. Hence, the single field winding assumption is reasonable for such generators. We label the single field winding with the letter $f$. The electrical diagram for the phase-$a$ stator winding is given in Figure \ref{phaseafig}.
In this diagram, $\lambda_a$ is the flux generated at the phase-$a$ winding, $r_a$ is the winding resistance, $V_a$ is the voltage at the terminals of the winding and $I_a$ is the current \textit{entering} through the positive pole of the winding terminal. The notation we choose for the current is called the \textit{motor notation}. One can obtain the \textit{generator notation} by replacing $I_a$ with $-I_a$. The diagram for the other phases ($b$ and $c$) and the field winding can be obtained by replacing the subscript $a$ in the diagram with the corresponding letters. From Kirchoff's Voltage Law, we have $$\dot{\lambda}_a = -r_a I_a + V_a$$ for the phase $a$ winding. The equations for the phases $b$ and $c$ can be obtained by replacing subscript $a$ with $b$ and $c$, respectively. Since the stator winding circuits are identical, we have $r = r_a = r_b = r_c$. We can write these equations in the vector form 
\begin{equation}
\dot{\lambda}_{abc} = -R I_{abc} + V_{abc},\label{eqn:elec1}
\end{equation} 
where $\lambda_{abc} = (\lambda_a, \lambda_b, \lambda_c, \lambda_f)$, $I_{abc} = (I_a, I_b, I_c, I_f)$, \mbox{$V_{abc} = (V_a, V_b, V_c, V_f)$} and \linebreak \mbox{$R = \textbf{diag}(r,r,r,r_f)$}.  In a synchronous generator with a single field winding, we can relate fluxes and currents using the equation \begin{equation}
\lambda_{abc} = \mathbb{L}_{abc} I_{abc},\label{fluxcur}
\end{equation} 
where
\begin{equation}
\mathbb{L}_{abc} = \begin{bmatrix} 
L_s + L_{s0} & -L_{s0} & -L_{s0} & L_{sf}\cos\left(\theta\right)\\
-L_{s0} & L_s + L_{s0} & -L_{s0} & L_{sf}\cos\left(\theta-\frac{2\pi}{3}\right)\\
-L_{s0} & -L_{s0} & L_s + L_{s0} & L_{sf}\cos\left(\theta+\frac{2\pi}{3}\right)\\
L_{sf}\cos\left(\theta\right) & L_{sf}\cos\left(\theta-\frac{2\pi}{3}\right) & L_{sf}\cos\left(\theta+\frac{2\pi}{3}\right) & L_{f}\\
\end{bmatrix}.\notag
\end{equation}
The inductance matrix $\mathbb{L}_{abc}$ is obtained from the inductance matrix in \cite[page 273]{FitzgeraldBook} by neglecting the saliency terms. We can define the total magnetic energy stored in the windings as
$$
H_{\text{magnetic}} = \frac{1}{2}\lambda_{abc}^T \mathbb{L}^{-1}_{abc} \lambda_{abc}
$$
and express the electrical equation \eqref{eqn:elec1} using $H_{\text{magnetic}}$ as
\begin{equation}
\dot{\lambda}_{abc} = -R \frac{\partial H_{\text{magnetic}}}{\partial \lambda_{abc}} + V_{abc} \label{eqn:elec2}.
\end{equation}
\subsection{Port-Hamiltonian Model of a Single Generator}
Using the total magnetic energy $H_{\text{magnetic}}$ defined in Section \ref{sec:sg:electrical}, we can explicitly compute the electrical torque $\tau_e$ in \eqref{eqn:mech2} as
$$
\tau_e = \frac{\partial H_{\text{magnetic}}}{\partial \theta}.
$$
The Hamiltonian for the single generator is the sum of the kinetic and magnetic energies, i.e.,
$$H = H_{\text{kinetic}} + H_{\text{magnetic}}.$$
Note that $H_{\text{kinetic}}$ does not depend on $\theta$ and $\lambda_{abc}$ and $H_{\text{magnetic}}$ does not depend on $M\omega$. Replacing the electrical torque expression in \eqref{eqn:mech2}, the equations \eqref{eqn:mech1},\eqref{eqn:mech2} and \eqref{eqn:elec1} can be written in the form
\begin{align}
\dot{\theta} &= \omega, \label{eqn:gen4} \\
M\dot{\omega} &= -D\frac{\partial H}{\partial (M\omega)} + \tau_m - \frac{\partial H}{\partial \theta}, \label{eqn:gen5}\\
\dot{\lambda}_{abc} &= -R \frac{\partial H}{\partial \lambda_{abc}} + V_{abc}.\label{eqn:gen6}
\end{align}
If we define the energy variables $\chi = (\theta,M\omega,\lambda_{abc})$ we obtain the port-Hamiltonian representation of equations \eqref{eqn:gen4}-\eqref{eqn:gen6} with state $\chi$, input $(\tau_m,V_{abc})$ and:
\begin{equation}
\mathcal{J}-\mathcal{R} = \begin{bmatrix}
0 & 1 & 0_{1 \times 4}\\
-1 & -D & 0_{1 \times 4}\\
0_{4 \times 1} & 0_{4 \times 1} & -R\\
\end{bmatrix},\quad g_{abc}=\begin{bmatrix}
0 & 0_{1 \times 4}\\
1 & 0_{1 \times 4}\\
0_{4 \times 1} & 1_{4}
\end{bmatrix}.\notag
\end{equation}

\subsection{Transformation from $abc$ Domain to $xyz$ Domain and a Simplifying Assumption}
Steady state currents and voltages for the $abc$ phases of the single generator are sinusoidal waveforms. In order to focus on the simpler problem of stability of equilibrium points, we perform a change of coordinates $T_\theta: \mathbb{R}^4 \rightarrow \mathbb{R}^4$ defined by the point-wise linear map
\begin{equation}
T_\theta =
\sqrt{\frac{2}{3}}\begin{bmatrix}
\cos(\theta)  &
\cos(\theta-\frac{2\pi}{3})  &
\cos(\theta+\frac{2\pi}{3}) & 0\\
\sin(\theta) &
\sin(\theta-\frac{2\pi}{3}) &
\sin(\theta+\frac{2\pi}{3}) & 0\\
\frac{\sqrt{2}}{2} & \frac{\sqrt{2}}{2} & \frac{\sqrt{2}}{2} & 0\\
0 & 0 & 0 & \sqrt{\frac{3}{2}}
\end{bmatrix}\label{transform}.
\end{equation}
with the inverse $T^{-1}_\theta = T^T_\theta$.
\begin{remark}
In the power systems literature, it is assumed that the generator rotor angles rotate with a speed that is very close to synchronous speed $\omega_s$, i.e., $\dot{\theta} \approx \omega_s$. If we integrate this approximation and assume zero initial conditions, we obtain $\theta = \omega_s t$. When we replace $\theta = \omega_s t$ in \eqref{transform}, the upper $3$-by-$3$ matrix of \eqref{transform} becomes a transformation that maps balanced waveforms with frequency $\omega_s$ to constant values, also known as \textit{Park's transformation} \cite{SauerBook}.
\end{remark}
Using \eqref{transform}, we can map $abc$-domain currents \mbox{$I_{abc} = (I_a, I_b, I_c, I_f)$} to $xyz$-domain currents $I_{xyz} = (I_x, I_y, I_z, I_f) = T_\theta I_{abc}$.  Note that the field winding current $I_f$ is not affected by the change of coordinates. We define $xyz$-winding voltages as $$V_{xyz} = (V_x, V_y, V_z, V_f) = T_\theta V_{abc},$$ and $xyz$-winding fluxes as $$\lambda_{xyz} = (\lambda_x, \lambda_y, \lambda_z, \lambda_f) = T_\theta \lambda_{abc},$$ in a similar fashion. We obtain the Hamiltonian $H$ in the new coordinates as
$$
H = \frac{1}{2}\lambda_{abc}^T \mathbb{L}^{-1}_{abc}\lambda_{abc}+ \frac{1}{2}M\omega^2 = \frac{1}{2}\lambda_{xyz}^T \mathbb{L}^{-1}_{xyz}\lambda_{xyz} + \frac{1}{2}M\omega^2,
$$
where $\mathbb{L}_{xyz} = \left( T_\theta\mathbb{L}^{-1}_{abc}T^{-1}_\theta\right)^{-1}$ is given by
\begin{equation}
\mathbb{L}_{xyz} = 
\begin{bmatrix}
L_s + 2L_{s0} & 0 & 0 & \sqrt{\frac{3}{2}}L_{sf}\\
0 & L_s + 2L_{s0} & 0 & 0\\
0 & 0 & L_s - L_{s0} & 0\\
\sqrt{\frac{3}{2}}L_{sf} & 0 & 0 & L_f
\end{bmatrix}.\label{eqn:matrixxyz}
\end{equation}
Equations \eqref{eqn:gen4}-\eqref{eqn:gen6} can be written in the $xyz$-domain as
\begin{align}
\dot{\theta} &= \omega, \label{eqn:phsgen4}\\
\dot{\xi} &= \left(\mathcal{J}(\xi) - \mathcal{R}\right)\frac{\partial H}{\partial \xi} + g\begin{bmatrix}\tau_m\\V_{xyz}\end{bmatrix}\label{eqn:phsgen5},
\end{align}
where $\xi = (M\omega, \lambda_{xyz})$ and
$$
\mathcal{J}(\xi) = \begin{bmatrix}
0 & \lambda_y & -\lambda_x & 0 & 0\\
-\lambda_y & 0 & 0 & 0 & 0\\
\lambda_x & 0 & 0 & 0 & 0\\
0 & 0 & 0 & 0 & 0\\
0 & 0 & 0 & 0 & 0\\
\end{bmatrix},\quad g=1_5.
$$

At the desired steady state operation, the fluxes $\lambda_{xyz}$ are constant and $\omega$ is the synchronous velocity $\omega_s$. Therefore, we can safely disregard \eqref{eqn:phsgen4} and focus on the stability of the equilibria of \eqref{eqn:phsgen5}. From the last row of \eqref{eqn:phsgen5}, we have
\begin{equation}
\dot{\lambda}_f = -r_f I_f + V_f\label{eqn:fieldwind}
\end{equation} 
which can be expressed in terms of currents as:
$$
L_f \dot{I}_f = -\sqrt{\frac{3}{2}}L_{sf} \dot{I}_x - r_f I_f + V_f
$$
by using the equality $\lambda_f = \sqrt{\frac{3}{2}}L_{sf} I_x + L_f I_f$ that follows from $\lambda_{xyz} = \mathbb{L}_{xyz} I_{xyz}$. Note that we can always design a control law acting on the field winding terminals by choosing the voltage $V_f$ according to $$V_f = \sqrt{\frac{3}{2}}L_{sf} \dot{I}_x + r_f I_f + \alpha\left(I_f - I^*_f\right)$$ for some $\alpha < 0$ and a constant reference value $I^*_f$. This controller keeps the field current constant and justifies the following assumption:

\begin{assumption}
The field winding current $I_f$ is constant.
\end{assumption}

If we use \eqref{fluxcur}, and consider the field winding current $I_f$ to be constant, we can express \eqref{eqn:phsgen5} in terms of currents:
\begin{align}
M \dot{\omega} &= -D \omega - L_m I_f I_y + \tau_m ,\label{eqn:curs1} \\
L_{ss}\dot{I}_x &= -r I_x - \omega L_{ss} I_y+ V_x, \label{eqn:curs2} \\
L_{ss}\dot{I}_y &= -r I_y  + \omega L_{ss}I_x + \omega L_m I_f+ V_y, \label{eqn:curs3} \\
(L_{ss}-3L_{s0})\dot{I}_z &= -r I_z + V_z,\label{eqn:curs4}
\end{align}
where $L_{ss} = L_s+2L_{s0}$ and $L_{m} = \sqrt{\frac{3}{2}}L_{sf}$. 

\subsection{Equilibria of a Single Generator}
\label{sec:stability:sg}
In this section, we study the equilibria of a single generator. Recall that sinusoidal waveforms  in the $abc$ coordinates are mapped to constant values on the $xyz$ coordinates. Therefore, equilibria of \eqref{eqn:curs1}-\eqref{eqn:curs4} are points rather than sinusoidal trajectories. We can find the equilibrium currents $I^*_x$, $I^*_y$, and $I^*_z$ that satisfy \eqref{eqn:curs2}-\eqref{eqn:curs4} when the voltages across the generator terminals, $V_x$, $V_y$, and $V_z$, are constant and equal to $V^*_x$, $V^*_y$, and $V^*_z$, respectively, by solving the algebraic equations
\begin{align}
0 &= -r I^*_x - \omega L_{ss} I^*_y+ V^*_x, \label{eqn:phasex} \\
0 &= -r I^*_y  + \omega L_{ss}I^*_x + \omega L_m I_f+ V^*_y, \label{eqn:phasey} \\
0 &= -r I^*_z + V^*_z,\label{eqn:phasez}
\end{align}
to obtain $I^*_z = \frac{V^*_z}{r}$ and
\begin{align}
I^*_{x} &= \frac{-\omega^2 L_m L_{ss} I_{f}-\omega L_{ss} V^*_y+rV^*_x}{r^2 + \omega^2 L^2_{ss}}  \label{xeq} \\
I^*_{y} &= \frac{\omega L_{ss}V^*_x+\omega r L_mI_{f}+rV^*_y}{r^2 + \omega^2 L^2_{ss}}. \label{yeq}
\end{align}
The values of $\omega$ are obtained by replacing \eqref{yeq} into the algebraic equation obtained by setting $\dot{\omega} = 0$ in \eqref{eqn:curs1}. This results in a third order polynomial equation in $\omega$. For any given $\omega_s \in \mathbb{R}$, if we choose
\begin{equation}
\tau^*_m = L_m I_f\left(\frac{\omega_s L_{ss} V^*_x+\omega_s rL_m I_{f}+rV^*_y}{r^2 + (\omega_s)^2 L^2_{ss}}\right) + D\omega_s,\label{eqn:torque}
\end{equation}
it is easy to show that one of the solutions of $\dot{\omega} = 0$ is $\omega = \omega_s$. Therefore, we can always choose a torque value $\tau_m$ such that for any given steady state inputs $(V^*_x,V^*_y,V^*_z)$ and desired synchronous velocity $\omega_s$,  one of the solutions of the equations \eqref{eqn:curs1}--\eqref{eqn:curs4} is
\begin{equation} 
(M\omega,L_{ss}I_x,L_{ss}I_y,L_{ss}I_z) = (M\omega_s,L_{ss}I^*_x,L_{ss}I^*_y,L_{ss}I^*_z),\label{eqn:eql}
\end{equation}
with $I^*_x$ and $I^*_y$ given by \eqref{xeq} and \eqref{yeq}, respectively and \mbox{$I^*_z = \frac{V^*_z}{r}$}. Note that, in addition to $\omega_s$, the equation $\dot{\omega} = 0$ has two other solutions. For each solution we potentially have an equilibrium point. Hence, in general we have three equilibrium points. By analyzing the coefficients of the polynomial equation $\dot{\omega}=0$ it is not difficult to show that the only real solution of $\dot{\omega}=0$ is $\omega_s$ iff
\begin{equation}
-4D^2 r^2 -4 D I_f L_m r \left(I_f L_m + L_{ss}I^*_x\right) + \left(I_f L_m L_{ss} I^*_y\right)^2 < 0, \label{bcond}
\end{equation}
where $I^*_x$ and $I^*_y$ are obtained by replacing $\omega$ with $\omega_s$ in \eqref{xeq} and \eqref{yeq}, respectively. Inequality \eqref{bcond} is a necessary condition for global asymptotic stability of the equilibrium $\xi^*$. In the next section, we obtain sufficient conditions by identifying constraints on the generator parameters that lead to a global Lyapunov function for the equilibrium $\xi^*$.

\subsection{Stability of  a Single Generator}
In this section, we provide sufficient conditions for the equilibrium point computed in Section \ref{sec:stability:sg} to be globally asymptotically stable. A natural choice for Lyapunov function candidate is the Hamiltonian $H$ of the single generator. However the minimum of $H$ occurs at the origin instead of $\xi^* = (M\omega_s, \lambda^*_{xyz})$, where $\lambda^*_{xyz} = \mathbb{L}_{xyz}I^*_{xyz}$. We shift the minimum of the Hamiltonian to $\xi^*$ by defining a function we call the \textit{shifted Hamiltonian}. Explicitly, the shifted Hamiltonian is given as
\begin{equation}
\hat{H} = \frac{1}{2}\left(\lambda_{xyz}-\lambda^*_{xyz}\right)^T \mathbb{L}^{-1}_{xyz}\left(\lambda_{xyz}-\lambda^*_{xyz}\right) + \frac{1}{2}M\left(\omega-\omega_s\right)^2.
\end{equation}
We also define the shifted state by $\hat{\xi} = \xi - \xi^*$. It is easy to check that we have
\begin{equation}
\frac{\partial \hat{H}}{\partial \hat{\xi}} = \frac{\partial H}{\partial \xi} - \frac{\partial H}{\partial \xi}\Big|_{\xi^*}.
\end{equation}
where $\frac{\partial H}{\partial \xi}\Big|_{\xi^*}$ is the gradient of the Hamiltonian $H$ with respect to $\xi$, evaluated at $\xi = \xi^*$. Note that $\hat{H}$ is positive definite and $\hat{H} = 0$ implies $\hat{\xi} = 0$, which in turn implies $\xi = \xi^*$. Therefore, in order to prove that $\xi^*$ is globally asymptotically stable, it is enough to show that $\frac{d\hat{H}}{dt} < 0$.  The time derivative of $\xi^*$ is given by
\begin{equation}
\frac{d}{dt}\xi^* = 0 = \left(\mathcal{J}(\xi^*)-\mathcal{R}\right)\frac{\partial H}{\partial \xi}\Big|_{\xi^*} + g\begin{bmatrix}\tau^*_m\\V^*_{xyz}\end{bmatrix},\label{eqn:phsgen7}
\end{equation}
From equation \eqref{eqn:phsgen5} and $V_{xyz} = V^*_{xyz}$, we obtain
\begin{align}
\dot{\xi} &= \dot{\hat{\xi}} = (\mathcal{J}(\xi) - \mathcal{R})\frac{\partial H}{\partial \xi} + g \begin{bmatrix}\tau^*_m\\V_{xyz}\end{bmatrix} \notag \\
&=(\mathcal{J}(\xi^*) + \mathcal{J}(\hat{\xi}) - \mathcal{R})\left(\frac{\partial \hat{H}}{\partial \hat{\xi}} + \frac{\partial
H}{\partial \xi}\Big\vert_{\xi^*}\right) + g \begin{bmatrix}\tau^*_m\\V_{xyz}\end{bmatrix}\notag \\
&=\left( (\mathcal{J}(\xi^*) - \mathcal{R}) \frac{\partial
H}{\partial \xi}\Big\vert_{\xi^*} +  g \begin{bmatrix}\tau^*_m\\V^*_{xyz}\end{bmatrix} \right) \notag \\
&+ \mathcal{J}(\hat{\xi}) \frac{\partial H}{\partial \xi}\Big\vert_{\xi^*} + (\mathcal{J}(\xi^*) + \mathcal{J}(\hat{\xi}) - \mathcal{R})\frac{\partial \hat{H}}{\partial \hat{\xi}} +  g\begin{bmatrix}0\\\hat{V}_{xyz}\end{bmatrix}.\label{phsingle2}
\end{align}
where $\hat{V}_{xyz} = V_{xyz}-V_{xyz}^*$ and we used the equality \mbox{$\mathcal{J}(\xi)=\mathcal{J}(\xi^*)+\mathcal{J}(\hat{\xi})$}. From \eqref{eqn:phsgen7}, we know that the term inside parentheses in \eqref{phsingle2} is equal to zero. Hence \eqref{phsingle2} implies 
\begin{equation}
\dot{\hat{\xi}} =  \mathcal{J}(\hat{\xi}) \frac{\partial
H}{\partial \xi}\Big\vert_{\xi^*} + (\mathcal{J}(\xi^*) + \mathcal{J}(\hat{\xi}) - \mathcal{R})\frac{\partial \hat{H}}{\partial \hat{\xi}} +  g\begin{bmatrix}0\\\hat{V}_{xyz}\end{bmatrix}.
\end{equation}
Taking the time derivative of the shifted Hamiltonian $\hat{H}$, we get
\begin{equation}
\frac{d \hat{H}}{dt} = \frac{\partial \hat{H}}{\partial \hat{\xi}}^T\dot{\hat{\xi}} = \frac{\partial \hat{H}}{\partial \hat{\xi}}^T\mathcal{J}(\hat{\xi})\frac{\partial
H}{\partial \xi}\Big\vert_{\xi^*} - \frac{\partial \hat{H}}{\partial \hat{\xi}}^T \mathcal{R}\frac{\partial \hat{H}}{\partial \hat{\xi}} + \hat{V}^T_{xyz}\hat{I}_{xyz} \label{eqn:hamildecay}
\end{equation}
where $\hat{I}_{xyz} = I_{xyz}-I_{xyz}^*$. Note that the last element of $\hat{I}_{xyz}$ is zero since a constant field winding current implies \mbox{$\hat{I}_f = I_f - I^*_f = 0$}. We can write the first term in the \mbox{right-hand} side of \eqref{eqn:hamildecay} as a quadratic function of $\frac{\partial \hat{H}}{\partial \hat{\xi}}$. Explicitly,
\begin{align}
&\frac{\partial \hat{H}}{\partial \hat{\xi}}^TJ(\hat{\xi})\frac{\partial
H}{\partial \xi}\Big\vert_{\xi^*} =\hat{\omega}\hat{\lambda}_yI^*_x - \hat{\omega}\hat{\lambda}_xI^*_y + \omega_s\left(\hat{\lambda}_x\hat{I}_y-\hat{\lambda}_y\hat{I}_x\right).\notag\\
& = \hat{\omega}\hat{I}_yL_{ss}I^*_x - \hat{\omega}\hat{I}_x L_{ss}I^*_y  \notag \\
&= \frac{\partial
\hat{H}}{\partial \hat\xi}^T\begin{bmatrix}0 & -\frac{1}{2}L_{ss} I^*_y & \frac{1}{2}L_{ss} I^*_x & 0 & 0\\
-\frac{1}{2}L_{ss} I^*_y  & 0 & 0 & 0 & 0 \\
\phantom{-}\frac{1}{2}L_{ss} I^*_x  & 0 & 0 & 0 & 0\\
0 & 0 & 0 & 0 & 0 \\
0 & 0 & 0 & 0 & 0
\end{bmatrix}\frac{\partial
\hat{H}}{\partial \hat\xi}\label{eqn:fquad2}.
\end{align}
where we used $\hat{\lambda}_{xyz} = \mathbb{L}_{xyz}\hat{I}_{xyz}$ to eliminate the flux variables. Replacing \eqref{eqn:fquad2} in \eqref{eqn:hamildecay}, we obtain
\begin{equation}
\frac{d\hat{H}}{dt} = \frac{\partial
\hat{H}}{\partial \hat\xi}^T\mathcal{P}\frac{\partial
\hat{H}}{\partial \hat\xi}+\hat{V}^T_{xyz}\hat{I}_{xyz}\label{eqn:decayrate},
\end{equation}
where
\begin{equation}
\mathcal{P} = \begin{bmatrix}-D & -\frac{1}{2}L_{ss} I^*_y & \frac{1}{2}L_{ss} I^*_x & 0 & 0\\
-\frac{1}{2}L_{ss} I^*_y  & -r & 0 & 0 & 0 \\
\phantom{-}\frac{1}{2}L_{ss} I^*_x  & 0 & -r & 0 & 0 \\
0 & 0 & 0 & -r & 0 \\
0 & 0 & 0 & 0 & -r_f
\end{bmatrix}.\label{eqn:pmatrix}
\end{equation}
The eigenvalues of the matrix $\mathcal{P}$ are $\lambda_1 = -r_f$, $\lambda_2 = \lambda_3 = -r$ and $$\lambda_{4,5} = -\frac{D+r}{2} \pm \frac{\sqrt{D^2 - 2Dr + r^2 + (L_{ss}I^*_x)^2 + (L_{ss}I^*_y)^2}}{2}.$$ 
Since we have $V_{xyz} = V^*_{xyz}$, i.e. $\hat{V}_{xyz} = 0$, if $\mathcal{P}$ is negative definite, then $\frac{d\hat{H}}{dt}<0$. It is easy to check that if 
\begin{equation}
(L_{ss}I^*_x)^2 + (L_{ss}I^*_y)^2 < 4 D r. \label{smcond2}
\end{equation}
holds, then $\lambda_{4,5} < 0$, which implies that the matrix $\mathcal{P}$ in \eqref{eqn:decayrate} is negative definite. Hence, if \eqref{smcond2} holds, we have $\frac{d\hat{H}}{dt}<0$, which in turn implies that $\xi^*$ is globally asymptotically stable. We can summarize the preceding discussion in the following result.

\begin{theorem}
\label{thm1}
Let $\xi^*$ be an equilibrium point of the single generator, described by equation \eqref{eqn:phsgen5} when we have \mbox{$\tau_m=\tau_m^*$} and $V_{xyz} = V^*_{xyz}$. The equilibrium point $\xi^*$ is globally asymptotically stable if 
\begin{equation}
(L_{ss}I^*_x)^2 + (L_{ss}I^*_y)^2 < 4 D r. \label{smcond}
\end{equation}
\end{theorem}

It is useful to express inequality~(\ref{smcond}) in terms of \mbox{$d$-axis} and \mbox{$q$-axis} currents. We know that the \mbox{$x$-axis} and \mbox{$y$-axis} currents are different from the traditional \mbox{$d$-axis} and \mbox{$q$-axis} currents during the transient stage. However, we have $$(I^*_x)^2 + (I^*_y)^2 = (I^*_d)^2 + (I^*_q)^2$$ at equilibrium. Thus, we can replace~(\ref{smcond}) with $(L_{ss}I^*_d)^2 + (L_{ss}I^*_q)^2 < 4 D r$. Note that we are using motor reference directions. In order to find the generator currents, we need to replace $I^*_x$ and $I^*_y$ by $-I^*_x$ and $-I^*_y$, respectively. However, this change in reference directions does not effect \eqref{smcond}. Condition \eqref{smcond} relates the total magnetic energy stored on the generator windings at steady state (left hand side of \eqref{smcond}) to the dissipation terms $D$ and $r$. This relation gives us a set of admissible steady-state currents in $xy$-coordinates (or alternatively, $dq$-coordinates) that lead to global asymptotical stability.

\begin{remark}
One can verify that if inequality \eqref{smcond} holds, then inequality \eqref{bcond} also holds while the converse is not true. This is to be excepted since global asymptotical stability requires a unique equilibrium.
\end{remark}

\begin{remark}
In the single machine infinite bus scenario, a generator is connected to an infinite bus modeling the power grid as a constant voltage source. The analysis of the single machine in this section, which is based on the assumption that the terminal voltages are constant, can also be seen as the analysis of a single machine connected to an infinite bus. In the classical analysis of this scenario \cite{AndersonBook}, there are multiple equilibrium points and energy based conditions for local stability are obtained. The analysis in this section shows that in fact a single equilibrium exists, under certain assumptions on the generator parameters, and that \textit{global} asymptotical stability is also possible. Such conclusions are not possible to obtain using the classical models as they are not detailed enough.
\end{remark}

\section{Stability Analysis\\of Multi-Machine Power Systems}
\subsection{Multi-Machine Power System Model}
\label{sec:mmmodel}
We consider a multi-machine power system consisting of $N$ generators, $M$ loads and a transmission grid connecting the generators and the loads. We distinguish between different generators by labeling each variable in the generator model with the subscript $i\in \{1,2,\hdots, N\}$. We make the following assumption about the multi-machine power system.

\begin{assumption}
The transmission network can be modeled by an asymptotically stable linear port-Hamiltonian system with Hamiltonian $H_\text{grid}$.\label{asm:network}
\end{assumption}

Concretely, this assumption states that whenever the inputs to the transmission network are zero, $H_\text{grid}$ can be used as a quadratic Lyapunov function proving global asymptotic stability of the origin. Although it may appear strong, we note that it holds in many cases of interest. In particular, it is satisfied whenever we use short or medium length approximate models to describe transmission lines in  arbitrary network topologies. Furthermore, we discuss in Remark~\ref{Rmk:Weak} how it can be relaxed.

We denote the three-phase voltages across the load terminals and currents entering into the load terminals by $V_{\ell,abc,j}$ and $I_{\ell,abc,j}$, respectively. Here, we use the letter $\ell$ to distinguish the currents and voltages that correspond to a load from the ones that correspond to a generator. The current entering into the load terminals when we set $V_{\ell,abc,j} = V^*_{\ell,abc,j}$ is denoted by $I^*_{\ell,abc,j}$. It follows from the linearity assumption on the transmission network that we can perform an affine change of coordinates so that in the new coordinates we have 
\begin{equation}
\hat{u}^T_\text{grid}\hat{y}_\text{grid} + \sum^N_{i=1}\hat{V}^T_{xyz,i}\hat{I}_{xyz,i} + \sum^M_{j=1}\hat{V}^T_{\ell,abc,j}\hat{I}_{\ell,abc,j} = 0\label{eqn:incpp}
\end{equation}
where $\hat{u}_\text{grid}$ and $\hat{y}_\text{grid}$ are the input and the output of the port-Hamiltonian model of the grid in the new coordinates with shifted Hamiltonian $\hat{H}_\text{grid}$, $\hat{V}_{\ell,abc,j} = V_{\ell,abc,j} - V^*_{\ell,abc,j}$, and $\hat{I}_{\ell,abc,j} = I_{\ell,abc,j} - I^*_{\ell,abc,j}$. Equation \eqref{eqn:incpp} represents an ``incremental power balance", i.e., a power balance in the shifted variables. Intuitively, it states that the net incremental power supplied by the generators and the loads is equal to the net incremental power received by the transmission grid.

\subsection{Load Models}
We make the following assumption regarding loads.

\begin{assumption}
Each load is described by one of the following models:
\begin{itemize}
\item a symmetric three-phase circuit with each phase being an asymptotically stable linear electric circuit;
\item a constant current source.
\end{itemize}
\label{asm:cload}
\end{assumption}

The proposed load models are quite simple and a subset of the models used in the power systems literature. It has recently been argued \cite{Raja2012} that the increase of DC loads, such as computers and appliances, interfacing the grid through power electronics intensifies the nonlinear character of the loads. However, there is no agreement on how such loads should be modeled. In fact, load modeling is still an area of research \cite{Milanovic2013}. The first class of models in Assumption~\ref{asm:cload} contains the well-known constant impedance model in the power systems literature. Constant impedance load models are commonly used in transient stability analysis \cite{FouadBook} and can be used to study the transient behavior of induction motors~\cite{FitzgeraldBook}. According to the IEEE Task Force on Load Representation for Dynamic Performance, more than half of the energy generated in the United States is consumed by induction motors \cite{IEEE1995}. This observation, together with the fact that these three-phase induction motors can be modeled as three-phase circuits with each phase being a series connection of a resistor, an inductor and a voltage drop justifies the constant impedance model usage in transient stability studies. The $RL$-circuit model suggested for induction motors in \cite{IEEE1995} is also captured by Assumption~\ref{asm:cload}. In \cite{IEEE1995}, it is also stated that lighting loads behave as resistors in certain operational regions. This observation also suggests the usage of constant impedances for modeling the aggregated behavior of loads. The second load model in Assumption~\ref{asm:cload} is also common in the power systems literature \cite{Milanovic2013}.

Any asymptotically stable linear electrical circuit has a unique equilibrium and admits a port-Hamiltonian representation with Hamiltonian $H_{\ell,abc,j}$. By performing a change of coordinates, we can obtain a port-Hamiltonian system for the shifted coordinates with the shifted Hamiltonian $\hat{H}_{\ell,abc,j}$ satisfying:
\begin{equation}
\frac{d\hat{H}_{\ell,j}}{dt} <  \hat{V}^T_{\ell,abc,j}\hat{I}_{\ell,abc,j}.\label{eqn:decayrateload}
\end{equation}
Let us now consider constant current loads. If a load $j$ draws constant current  from the network we have $I_{\ell,abc,j} = I^*_{\ell,abc,j}$. This implies $\hat{I}_{\ell,abc,j} = 0$ and the contribution of the constant current load $j$ to the incremental power balance \eqref{eqn:incpp} is zero. This observation shows that we can neglect constant current loads since they do not contribute to the incremental power balance. Therefore, in the remainder of the paper we only consider the first type of loads in Assumption~\ref{asm:cload}.

\subsection{Stability of Multi-Machine Power Systems}
Let $\hat{H}_i$ be the shifted Hamiltonian for generator $i$ with respect to the equilibrium point \mbox{$\xi^*_i = (M_i\omega_s,\lambda^*_{xyz,i})$}, as defined in Section \ref{sec:stability:sg}. From Section \ref{sec:stability:sg}, we know that for every generator $i$ we have
\begin{equation}
\frac{d\hat{H}_i}{dt} = \frac{\partial \hat{H}_i}{\partial \hat{\xi}_i}^T \mathcal{P}_i \frac{\partial \hat{H}_i}{\partial \hat{\xi}_i}+ \hat{V}^T_{xyz,i}\hat{I}_{xyz,i},\label{eqn:decayratei}
\end{equation}
where $\mathcal{P}_i$ is a matrix obtained by adding subscript $i$ to the elements of the matrix $\mathcal{P}$ given by \eqref{eqn:pmatrix}. Using the definitions above, we select our candidate Lyapunov function as 
$$
\hat{H}_\text{total} = \hat{H}_\text{grid} + \sum^N_{i=1}\hat{H}_i + \sum^{M}_{j=1}\hat{H}_{\ell,j},
$$
where $\hat{H}_\text{grid}$ is the shifted Hamiltonian of the transmission grid that was introduced in Assumption \ref{asm:network}, Section \ref{sec:mmmodel}. Our objective is to show that the equilibrium $\mbox{$\xi^* = (\xi^*_1,\ldots,\xi^*_N)$}$ for the generators is globally asymptotically stable. Note that every equilibrium $\xi_i^*$ shares the same synchronous velocity $\omega_s$. Hence, asymptotical stability of $\xi^*$ implies that all the generators converge to the synchronous velocity $\omega_s$. In addition to synchronize the generators' angular velocity we also need to ensure that the currents flowing through the transmission network converge to preset values respecting several operational constraints such as thermal limits of the transmission lines. This will also be a consequence of asymptotical stability of the equilibrium $\xi^*$. When this equilibrium is reached, the voltages and currents at the generator terminals are $V^*_{abc,i}$ and $I^*_{abc,i}$, respectively. If we now regard the transmission network and the loads as being described by an asymptotically stable linear system driven by the inputs $V^*_{abc,i}$ and $I^*_{abc,i}$, we realize that all the voltages and currents in the transmission network and loads will converge to a unique steady state. We assume that such steady state, uniquely defined by $V^*_{abc,i}$ and $I^*_{abc,i}$, satisfies all the operational constraints.

Taking the time derivative of $\hat{H}_\text{total}$, we obtain 
\begin{align}
\frac{d\hat{H}_\text{total}}{dt} &= \frac{d\hat{H}_\text{grid}}{dt} + \sum^N_{i=1}\frac{d\hat{H}_i}{dt} + \sum^{M}_{j=1}\frac{d\hat{H}_{\ell,j}}{dt} \\
&\le \hat{u}^T_\text{grid}\hat{y}_\text{grid} +\sum^N_{i=1}\frac{d\hat{H}_i}{dt} + \sum^{M}_{j=1}\frac{d\hat{H}_{\ell,j}}{dt} \label{ngenmload1} \\
&= \sum^N_{i=1}\frac{\partial \hat{H}_i}{\partial \hat{\xi}_i}^T \mathcal{P}_i \frac{\partial \hat{H}_i}{\partial \hat{\xi}_i}, \label{ngenmload}
\end{align}
where \eqref{ngenmload1} follows from \eqref{eqn:passivity}, and \eqref{ngenmload} follows from \eqref{eqn:incpp}, \eqref{eqn:decayrateload}, and \eqref{eqn:decayratei}. If
\begin{equation}
(L_{ss,i} I^*_{x,i})^2 + (L_{ss,i} I^*_{y,i})^2 < 4 D_i r_i.\label{mmconservativer}
\end{equation}
holds for $i \in \{1,\ldots,N\}$, then $\mathcal{P}_i < 0$ for every $ i \in \{1,\ldots,N\}$ by Theorem \ref{thm1}. Therefore, if \eqref{mmconservativer} holds for every $ i \in \{1,\ldots,N\}$, we conclude from \eqref{ngenmload} that
\begin{equation}
\frac{d\hat{H}_\text{total}}{dt} \le \sum^N_{i=1}\frac{\partial \hat{H}_i}{\partial \hat{\xi}_i}^T \mathcal{P}_i \frac{\partial \hat{H}_i}{\partial \hat{\xi}_i} < 0.\label{eqn:negsemdef}
\end{equation}
This only shows that $\frac{d\hat{H}_\text{total}}{dt}$ is negative semi-definite. Since all of the Hamiltonians that constitute the total Hamiltonian $\hat{H}_\text{total}$ have compact level sets, the level sets of $\hat{H}_\text{total}$ are also compact. Hence, we can apply La Salle's Invariance Principle to conclude that all the trajectories converge to the largest invariant set contained in the set defined by
\begin{equation}
\sum^N_{i=1}\frac{\partial \hat{H}_i}{\partial \hat{\xi}_i}^T \mathcal{P}_i \frac{\partial \hat{H}_i}{\partial \hat{\xi}_i} = 0.\label{eqn:linvset}
\end{equation}
The left hand side of~\eqref{eqn:linvset} is a sum of negative definite quadratic terms (recall that $\mathcal{P}_i<0$) and thus only zero when $\frac{\partial \hat{H}_i}{\partial \hat{\xi}_i} = 0$ for all $i$. This implies $\xi=\xi^*$, hence the generator states globally asymptotically converge to $\xi^*$ if \eqref{eqn:negsemdef} holds. The preceding discussion is summarized in the following result.

\begin{theorem}
\label{thm2}
Consider a multi-machine power system with $N$ generators described by equations \eqref{eqn:phsgen5} with $\tau_{m,i} = \tau^*_{m,i}$, and $M$ loads satisfying Assumption \ref{asm:cload} interconnected by a transmission network satisfying Assumption \ref{asm:network}. Let $\xi^*$ be an equilibrium point for the generators that is consistent with all the equations describing the power system. The equilibrium $\xi^*$ is globally asymptotically stable if 
\begin{equation}
(L_{ss,i} I^*_{x,i})^2 + (L_{ss,i} I^*_{y,i})^2 < 4 D_i r_i. \label{mmconservative}
\end{equation}
holds for all $i \in \{1,\ldots,N\}$.
\end{theorem}

Theorem \ref{thm2} states that in order to check the stability of the multi-machine system, we only need to check a simple condition for each generator in the system. This makes our result compositional in the sense that the complexity of condition \eqref{mmconservative} is independent of the size of the network. All these conditions are bound together by $I_x^*$ and $I_y^*$ that obviously depend on the  whole network. However, the computation of the desired steady state currents needs to be performed for reasons other than transient stability and thus are assumed to be readily available.

\begin{remark}\label{Rmk:Weak}
We note that if Assumption~\ref{asm:network} is weakened from asymptotic stability to stability of the transmission network, the equilibrium $\xi^*$ is still globally asymptotically stable. However, the voltages and currents in the transmission network are no longer uniquely determined and may violate the operational constraints.
\end{remark}

Inequality \eqref{mmconservative} is a sufficient condition for asymptotic stability. Typically, the stator winding resistance $r_i$ for each generator $i$ is small and inequality~\eqref{mmconservative} is only satisfied for small steady state currents. However, inequality \eqref{mmconservative} can be enforced by actively controlling the voltage at the generator terminals using a static synchronous series compensator (SSSC), a FACTS device that is typically used for series compensation \cite{HingoraniBook} of real and reactive power. Using a SSSC we can introduce voltage drops of $R_i \left(I_{x,i} - I^*_{x,i}\right)$, $R_i \left(I_{y,i} - I^*_{y,i}\right)$, and $R_i \left(I_{z,i} - I^*_{z,i}\right)$ at the generator terminals without altering the current. The turn-on and turn-off times for the thyristors in a SCCC are at the level of microseconds~\cite{HingoraniBook}, small enough to enforce a voltage drop that is a piece-wise constant approximation of $R_i \left(I_{x,i} - I^*_{x,i}\right)$, $R_i \left(I_{y,i} - I^*_{y,i}\right)$, and $R_i \left(I_{z,i} - I^*_{z,i}\right)$. The approximation error can always be reduced by increasing the number of converter valves in the SSSC. By repeating the stability analysis in this section, while taking into consideration this new voltage drop, we arrive at the relaxed condition for global asymptotic stability:
\begin{equation}
(L_{ss,i} I^*_{x,i})^2 + (L_{ss,i} I^*_{y,i})^2 < 4 D_i \left(r_i + R_i\right).\label{mmconservative2}
\end{equation}
Since the power throughput of FACTS devices is in the order of megawatts \cite{HingoraniBook} we can choose a value for $R_i$ that is several order of magnitude larger than $r_i$. Therefore, the relaxed inequality~(\ref{mmconservative2}) allows for large steady-state currents and is  widely applicable to realistic examples.

\section{Simulation}
\begin{figure}[h!]
\centering
\begin{circuitikz}[scale = 0.75, transform shape]\draw
  (0,1.5) node[ground] {}
  (0,1.5) to [sI=$g_1$] (0,3)
  (6,1.5) node[ground] {}
  (6,1.5) to [sI=$g_2$] (6,3)
  (6,3) to [short, -*] (6,3)
  (0,3) to [TL=$z_1$,-*] (3,3)
  (3,3) to [TL=$z_2$,-*] (6,3)
  (3,3) to [european resistor,l_=$z_\ell$] (3,1.5)
  (3,1.5) node[ground] {};
\end{circuitikz}
\caption{Two-generator single-load scenario}
\label{mmdiagram}
\end{figure}
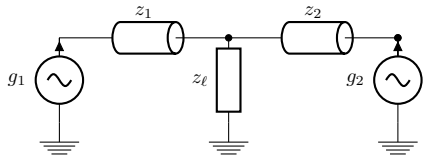

In this section we apply our results to the two-generator single-load scenario depicted in Figure~\ref{mmdiagram}.
The generators are connected to the load via transmission lines with impedances $z_1 = z_2 = \left(5 + j 0.1\omega_s\right)\Omega$ . The load impedance is $z_\ell = 1 \text{k}\Omega$. We use the generator parameters provided in \cite[Table 7.3]{SauerBook}. Using the provided $H_i$ values, the damping coefficients are selected as $D_1 = \frac{0.2H_1}{\omega_s} = 1.25$ MVAs and \mbox{$D_2 =  \frac{0.4H_2}{\omega_s}=0.68$}  MVAs as was done in \cite[Example 7.1]{SauerBook}. The stator winding resistances for the generators are taken to be $r_1 = r_2 = 0.05\Omega$. Using the values for $X_{d,i}$ and $X_{q,i}$  provided in \cite[Table 7.3]{SauerBook}, we obtain $L_{s,1} = 0.2049$ H and \mbox{$L_{s,2} = 1.2570$ H} from the equations $X_{d,i} = \omega_s L_{s,i}$. Since the parameters $L_{s0,i}$ and $L_{m,i}$ cannot be obtained from \cite[Table 7.3]{SauerBook} we assume \mbox{$L_{s0,i} = 0$} and \mbox{$L_{m,i} = L_{s,i}$}. The steady state phase-$xyz$ voltages are $V^*_{x,1} = -17.56$ kV, \mbox{$V^*_{y,1} = 280.16$ kV}, $V^*_{x,2} = -24.14$ kV, and $V^*_{y,2} = 278.76$ kV.
The steady state phase-$xyz$ currents satisfying the circuit constraints are $I^*_{x,1} = 19.83$ A, $I^*_{y,1} = -227.33$ A, \mbox{$I^*_{x,2} = 6.2$ A}, and $I^*_{y,2} = -50.9402$ A. The mechanical torques and field winding currents are selected so as to be consistent with these steady-state values. Inequality \eqref{bcond} reduces to $I^2_f L^2_m L^2_{ss} I^*_y < 0$ if the stator winding resistance $r$ is zero. Since the $y$-axis steady state current is negative for both generators,~(\ref{bcond}) is satisfied and the equilibrium is unique. We now investigate global asymptotic stability for this example. Condition \eqref{mmconservative} does not hold since the winding resistances for the generators are zero and this leads to \mbox{$(L_{ss,i} I^*_{x,i})^2 + (L_{ss,i} I^*_{y,i})^2 < 0 = 4D_i r_i$} for $i \in \{1,2\}$. In order to use the relaxed condition \eqref{mmconservative2}, we connect a static synchronous series compensator (SCCC) in series with the generator terminals providing the voltage drops $R_i\left(I_{x,i}-I^*_{x,i}\right)$, $R_i\left(I_{y,i}-I^*_{y,i}\right)$, and $R_i\left(I_{z,i}-I^*_{z,i}\right)$ in phases $x$, $y$, and $z$, respectively. Condition \eqref{mmconservative2} holds for generator $i$ if: 
$$R_i > \frac{(L_{ss,i} I^*_{x,i})^2 + (L_{ss,i} I^*_{y,i})^2}{4D_i}.$$ Replacing the generator parameters into this inequality, we obtain $R_1 > 0.437$ m$\Omega$ and $R_2 > 1.53$ m$\Omega$. We choose \mbox{$R_1 = R_2 = 10\Omega$} to satisfy these inequalities and provide enough damping.

\begin{figure}[!ht]
\centering
\includegraphics[scale=0.3]{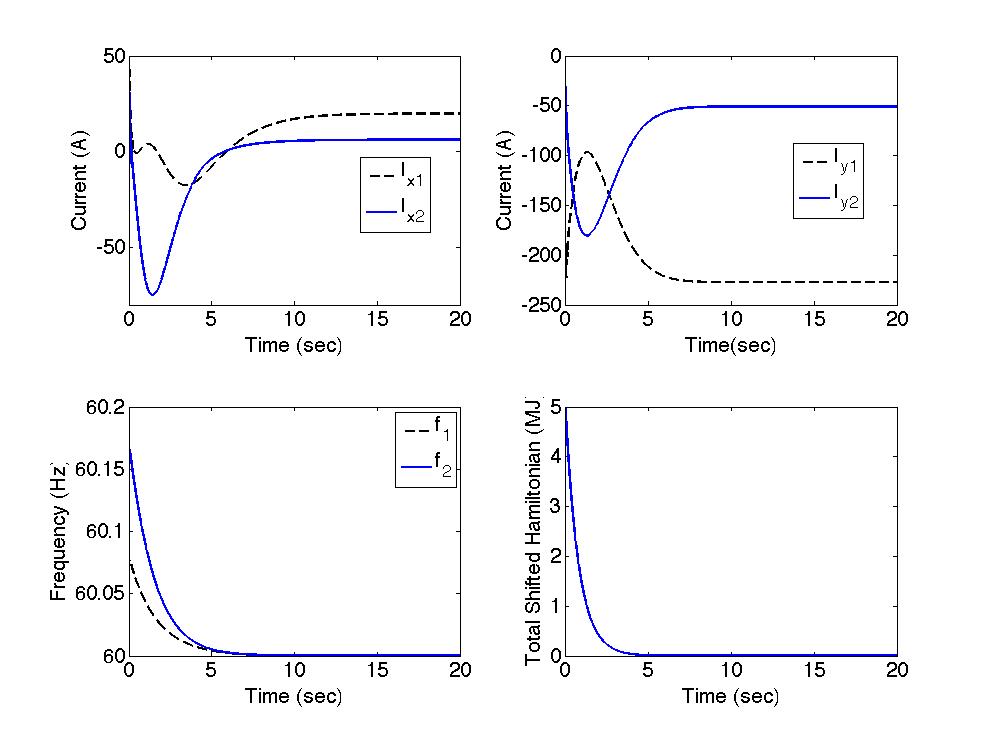}
\caption{Evolution of the generators' states ($x$-axis currents on upper left, $y$-axis currents on upper right, frequencies on lower left) and the value of the total shifted Hamiltonian (on lower right).}
\label{fig_sim}
\end{figure}

We numerically simulated the dynamics of the circuit in Figure~\ref{mmdiagram} to obtain the transient behavior following the occurrence of a fault. Without conjecturing anything about the nature of the fault or the pre-fault circuit, we simply assumed that the initial condition for the frequency of the generators lies in the set $[59.8, 60.2]$. For a generator current with steady state value $I^*$, we assumed that the initial condition for the current lies in the set \mbox{$[I^*-50, I^*+50]$}. With this assumption about the initial states, we performed numerical simulations for 25 randomly chosen initial state vectors. These simulations indicate that the generator states converge to the steady-state values as expected. We present in Figure~\ref{fig_sim} a typical trajectory corresponding to initial conditions \mbox{$I^*_{x,1}(0) = -62.46$} A, \mbox{$I^*_{x,2}(0) = -46.95$} A, $I^*_{y,1}(0) = -249.61$ A, \mbox{$I^*_{y,2}(0)=-70.71$ A}, $\omega_1(0) = 2\pi(60.08)$ rad$/$s, and \mbox{$\omega_2(0) = 2\pi(60.17)$} rad$/$sec. The reader can appreciate how the states and the value of the total shifted Hamiltonian converge to the desired values.

\section{Conclusion and Future Work}
This paper shows that transient stability analysis can be performed without using the hard-to-justify assumptions described in Section \ref{intro} and found in the classical literature on power systems. Instead, we employed first-principles models and obtained sufficient conditions for global transient stability that are applicable to networks with lossy transmission lines. Moreover, the proposed sufficient conditions for transient stability are compositional, i.e., we only need to check that each generator satisfies a simple inequality relating the steady state currents to the generators mechanical and electrical dissipation. Such test is far less expensive than contingency analysis based on numerical simulations.

Although transient stability is critical, equally important is a careful analysis of transients to ensure that operational limits are never violated. Such study is a natural next step in our investigations. Another direction for further research is how a careful modeling of transmission networks and loads can contribute to a refined transient stability analysis. In what regards the design of controllers, much is to be done on combining the use of FACTS devices with carefully designed excitation controllers to improve transient performance.

\end{document}